\documentclass[
reprint,
superscriptaddress,
amsmath,amssymb,
aps,
pra,
longbibliography
]{revtex4-1}

\usepackage{graphicx}
\usepackage{dcolumn}
\usepackage{bm}
\usepackage{hyperref}
\usepackage{siunitx} 
\usepackage[normalem]{ulem}
\usepackage{graphicx}
\usepackage{dcolumn}
\usepackage[utf8]{inputenc}
\usepackage{bm}
\usepackage{float}
\usepackage{url}

\usepackage{color}

\definecolor{darkgreen}{rgb}{0,0.6,0}

\graphicspath{{figures/}}

\begin{document}

\title{Radio-frequency methods for Majorana-based quantum devices: \\ fast charge sensing and phase diagram mapping}

\author{Davydas~Razmadze}
\thanks{These authors contributed equally to this work}
\affiliation{Center for Quantum Devices, Niels Bohr Institute, University of Copenhagen and Microsoft Quantum Lab Copenhagen, Universitetsparken 5, 2100 Copenhagen, Denmark}

\author{Deividas~Sabonis}
\thanks{These authors contributed equally to this work}
\affiliation{Center for Quantum Devices, Niels Bohr Institute, University of Copenhagen and Microsoft Quantum Lab Copenhagen, Universitetsparken 5, 2100 Copenhagen, Denmark}

\author{Filip~K.~Malinowski}
\affiliation{Center for Quantum Devices, Niels Bohr Institute, University of Copenhagen and Microsoft Quantum Lab Copenhagen, Universitetsparken 5, 2100 Copenhagen, Denmark}

\author{Gerbold~C.~M\'enard}
\affiliation{Center for Quantum Devices, Niels Bohr Institute, University of Copenhagen and Microsoft Quantum Lab Copenhagen, Universitetsparken 5, 2100 Copenhagen, Denmark}

\author{Sebastian~Pauka}
\affiliation{Center for Quantum Devices, Niels Bohr Institute, University of Copenhagen and Microsoft Quantum Lab Copenhagen, Universitetsparken 5, 2100 Copenhagen, Denmark}
\affiliation{ARC Centre of Excellence for Engineered Quantum Systems, School of Physics, The University of Sydney,  NSW 2006, Sydney Australia}

\author{Hung~Nguyen}
\affiliation{Center for Quantum Devices, Niels Bohr Institute, University of Copenhagen and Microsoft Quantum Lab Copenhagen, Universitetsparken 5, 2100 Copenhagen, Denmark}
\affiliation{Nano and Energy Center, Hanoi University of Science, VNU 120401, Hanoi, Vietnam}

\author{David~M.~T.~van~Zanten}
\affiliation{Center for Quantum Devices, Niels Bohr Institute, University of Copenhagen and Microsoft Quantum Lab Copenhagen, Universitetsparken 5, 2100 Copenhagen, Denmark}

\author{Eoin~C.~T.~O'Farrell}
\affiliation{Center for Quantum Devices, Niels Bohr Institute, University of Copenhagen and Microsoft Quantum Lab Copenhagen, Universitetsparken 5, 2100 Copenhagen, Denmark}

\author{Judith~Suter}
\affiliation{Center for Quantum Devices, Niels Bohr Institute, University of Copenhagen and Microsoft Quantum Lab Copenhagen, Universitetsparken 5, 2100 Copenhagen, Denmark}

\author{Peter Krogstrup}
\affiliation{Microsoft Quantum Materials Lab and Center for Quantum Devices, Niels Bohr Institute, University of Copenhagen, Kanalvej 7, 2800 Kongens Lyngby, Denmark}

\author{Ferdinand~Kuemmeth}
\affiliation{Center for Quantum Devices, Niels Bohr Institute, University of Copenhagen and Microsoft Quantum Lab Copenhagen, Universitetsparken 5, 2100 Copenhagen, Denmark}

\author{Charles~M.~Marcus}
\affiliation{Center for Quantum Devices, Niels Bohr Institute, University of Copenhagen and Microsoft Quantum Lab Copenhagen, Universitetsparken 5, 2100 Copenhagen, Denmark}


\begin{abstract}

Radio-frequency (RF) reflectometry is implemented in hybrid semiconductor-superconductor nanowire systems designed to probe Majorana zero modes. Two approaches are presented. In the first, hybrid nanowire-based devices are part of a resonant circuit, allowing conductance to be measured as a function of several gate voltages $\sim$40 times faster than using conventional low-frequency lock-in methods. In the second, nanowire devices are capacitively coupled to a nearby RF single-electron transistor made from a separate nanowire, allowing RF detection of charge, including charge-only measurement of the crossover from $2e$ inter-island charge transitions at zero magnetic field to $1e$ transitions at axial magnetic fields above 0.6 T, where a topological state is expected. Single-electron sensing yields signal-to-noise exceeding 3 and visibility 99.8$\%$ for a measurement time of 1~$\mu$s.

\end{abstract}
\vfill
\maketitle
\setlength{\parskip}{0.12cm plus4mm minus3mm}
\begin{figure}[tbh]
	\includegraphics[width=0.48\textwidth]{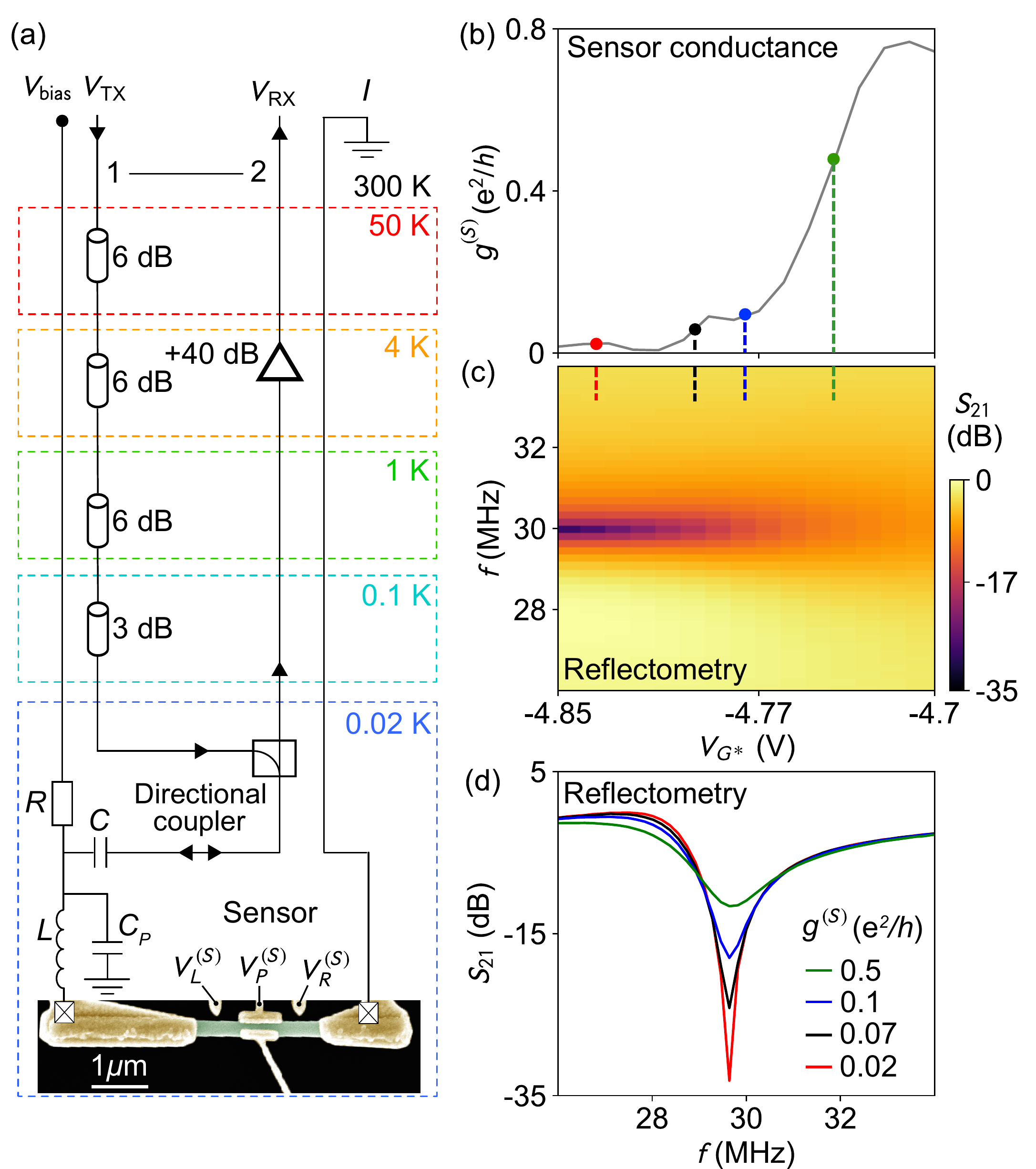}
	\caption{{\bf RF charge-sensing set-up}: (a) Circuit diagram of a nanowire (sensor) embedded in a resonant circuit allowing conductance (by measuring current $I(V_\mathrm{bias})$) or reflectometry measurement (by measuring reflected signal $V_\mathrm{RX}$), respectively (see main text). (b) Sensor conductance, $g^{(S)}$, as a function of sensor gate voltage, $V_{G^{*}}$ = $V_{L}^{(S)}$ = $V_{P}^{(S)}$ = $V_{R}^{(S)}$. (c) Scattering parameter, $S_{21}$, as a function of carrier frequency $f$, and $V_{G^{*}}$ acquired simultaneously with (b). $S_{21}(f)$ develops a dip at $f_{\rm res}$ $\sim$ 30 MHz, indicating that the matching condition of the resonator is approached towards low sensor conductance. (d) Vertical cuts of (c) for gate voltages indicated in (b). The on-resonance reflectometry signal acts as an alternative measure of $g^{(S)}$.}
\end{figure}

\section{INTRODUCTION}

Solid state quantum computation schemes that involve repeated measurement and feedback, including topological schemes \cite{AasenPRX16, Vijay, Plugge, Karzig} with potentially long coherence times \cite{Non-Abelian, Alicea}, nonetheless require fast readout of charge or current in order to operate on reasonable time scales \cite{Reiher7555}. For topological qubits based on Majorana modes in nanowires (NWs) with proximity-induced superconductivity, quasiparticle poisoning of Majorana modes constrain readout times to microseconds or faster \cite{Lossqpp}, which has already been demonstrated for superconducting \cite{SCq3,SCq2,SCq1,SCq4} and spin qubits \cite{Reilley1,spin1,petta2,Vandersypen}. 

Here, we report the realization of radio-frequency (RF) reflectometry in various configurations of InAs nanowires (NWs) with epitaxial Al, fabricated to form single or coupled Majorana islands, with and without proximal NW charge sensors. Device geometries were inspired by recent theoretical proposals for demonstrating elementary topological qubit operations in these systems \cite{AasenPRX16, Vijay, Plugge, Karzig}. Two approaches to fast measurements were investigated in detail. In the first, a resonator made from a cryogenic inductor and capacitor was coupled directly to the leads of the device \cite{karl1,karl2,karl3}, providing conductance measurement similar to what is obtained with a low-frequency (LF) lock-in amplifier, though considerably faster. In the second, a similar resonator was capacitively coupled to a proximal NW charge sensor configured for both LF and RF charge readout. The overall charge sensitivity was investigated as a function of measurement time, and found to yield a signal-to-noise ratio (SNR) for single-charge detection exceeding 3 and visibility of 99.8\% for an integration time of 1 $\mu$s, and correspondingly higher for longer integration times. Proximal NW charge sensors were found to be compatible with magnetic fields exceeding 1 T, the range needed to reach the topological regime \cite{Mourik,MT1,Quantized_Majorana,AlbretchNature}. All measurements were carried out in a dilution refrigerator (Oxford Instruments Triton 400) with base temperature $\sim$ 20 mK, equipped with a 6-1-1~T vector magnet. 

\section{Experimental setup}

Reflectometry signal is optimized by matching the circuit impedance $Z$, including device resistance $R_{\rm dev}$ to the characteristic impedance of the transmission line, $Z_0 = 50\,\Omega$. Near matching, the reflection coefficient of the resonant circuit, ($Z-Z_{0}$)/($Z+Z_{0}$), is sensitive to small changes in $R_{\rm dev}$ \cite{qpc_rf,impedance_matching}. To enable multiple simultaneous measurements, four RF resonant circuits with different discrete inductances in the range $L = 1.2$ - 4.7~$\mu$H, were coupled to a single directional coupler via coupling capacitor, \textit{C}. One such resonant circuit is depicted in Fig.~1(a). It consists of a ceramic-core chip inductor, parasitic capacitance, $C_P$, from bond wires and on-chip metal electrodes, and the device, with $R_{\rm dev}$ tuned by gate voltages. Parasitic capacitance was found to be unchanged over several cool-downs.  

LF lock-in measurements of differential conductance $g$ = d$I/$d$V|_{V_{\rm bias}}$ of either the device or the sensor were carried out in a two-wire voltage-bias configuration using a transimpedance (current to voltage) amplifier \cite{amp} connected to the drain of the device, providing voltage input to a lock-in amplifier (Stanford Research SR830). The voltage bias consisted of a DC component, $V_{\rm bias}$, and a LF component in the range of 4 - 10 $\mu$V at frequencies below 200 Hz.

Reflectometry measurements of either the device or the sensor were performed as follows. An RF carrier at frequency $f$ with amplitude $V_\mathrm{TX}$ was applied to the source lead following a series of attenuators at various temperature stages [Fig.~1(a)], giving a total of 21 dB of attenuation, with an additional 15 dB of attenuation from the directional coupler, mounted below the mixing chamber plate. After reflection from the device, the signal passed back through the directional coupler into a cryogenic amplifier (Caltech CITLF3; noise temperature $T_{n}$ =  4 K from 10 MHz to 2 GHz) with +40 dB of gain.  The output signal, $V_\mathrm{RX}$, was then detected using one of three methods: (1) using a network analyzer to measure $S_{21}\equiv 20 \log (V_\mathrm{RX}/V_\mathrm{TX})$ [Fig.~1(c)]; (2) using discrete analog components to demodulate by standard homodyne detection, followed by a fast-sampling oscilloscope (see Appendix \ref{appB} for details); (3) using an RF lock-in amplifier (Zurich Instruments UHFLI \cite{zi}). Each method has its advantages. Method (1) is convenient for quickly determining  if a change in device resistance has an effect on circuit impedance, which shows up as a change in the magnitude of $S_{21}$. Method (2) provides fast acquisition of phase maps at different gate configurations, particularly if the device is tuned into the regime of small charging energies. For these applications, Methods (2) and (3) are comparable. Method (3) has advantages in simultaneously measuring phase and magnitude of the reflected signal, and was used to quantify SNR of the proximal NW sensors and to detect charge occupancy of Majorana islands tuned to low barrier transmission. 

Figures 1(b-d) show a comparison of LF lock-in measurement and reflectometry measurement, $S_{21}(f)$, of conductance $g^{(S)}$ of a charge sensor as it is pinched off using electrostatic gates.  In the reflectometry measurement, $V_{RX}$ varies rapidly near the resonance frequency $f_{\rm res} \sim$ 30 MHz, yielding a dip in $S_{21}(f)$ that depends on the common gate voltage.  Line cuts of $S_{21}$ at different $V_{G^{*}}$ values are shown in Fig.~1(d). 
The depth of the resonance changed by approximately 21 dB as the sensor conductance, $g^{(S)}$, was decreased from 0.5 $e^{2}/h$ to 0.02 $e^{2}/h$. In this case, an increasing $R_{\rm dev}$ moves the resonator impedance toward matching.

\section{Conductance: LF lock-in versus RF reflectometry} 

Figure 2(a) shows a hybrid InAs/Al island (Device A) defined by Ti/Au gates that wrap around the NW, isolated by HfO$_{2}$ dielectric. Gate voltages $V_{L}$ and $V_{R}$ control coupling of the island to the leads, while three additional gates tuned the chemical potential and density on different parts of the island (see Appendix \ref{AppD}). Only the gate marked $V_{LP}$ in Fig.~2(a) was used, with the others fixed at zero volts. DC voltage $V_\mathrm{bias}$ was applied to the left lead while the right lead was connected to the RF circuit ($L$ = 2.7 $\mu$H, $f_{\rm res}$ $\sim$ 52 MHz) using method (2), described above. Simultaneous LF and RF measurements of pinch-off characteristic of the right barrier with left barrier fixed at $V_{L} \sim 1$~V is shown in Fig.~2(b). At positive $V_{R}$, the right side of the island was open and showed positive demodulated voltage $V_{\rm rf}$, while at negative  $V_{R}$, the right junction is closed and no current could flow through the island. Overall, $V_{\rm rf}$ was found to be proportional to $g$ measured with an LF lock-in, as shown in the inset of Fig.~2(b). 
 
\begin{figure}[tbh]
	\includegraphics[width=0.48\textwidth]{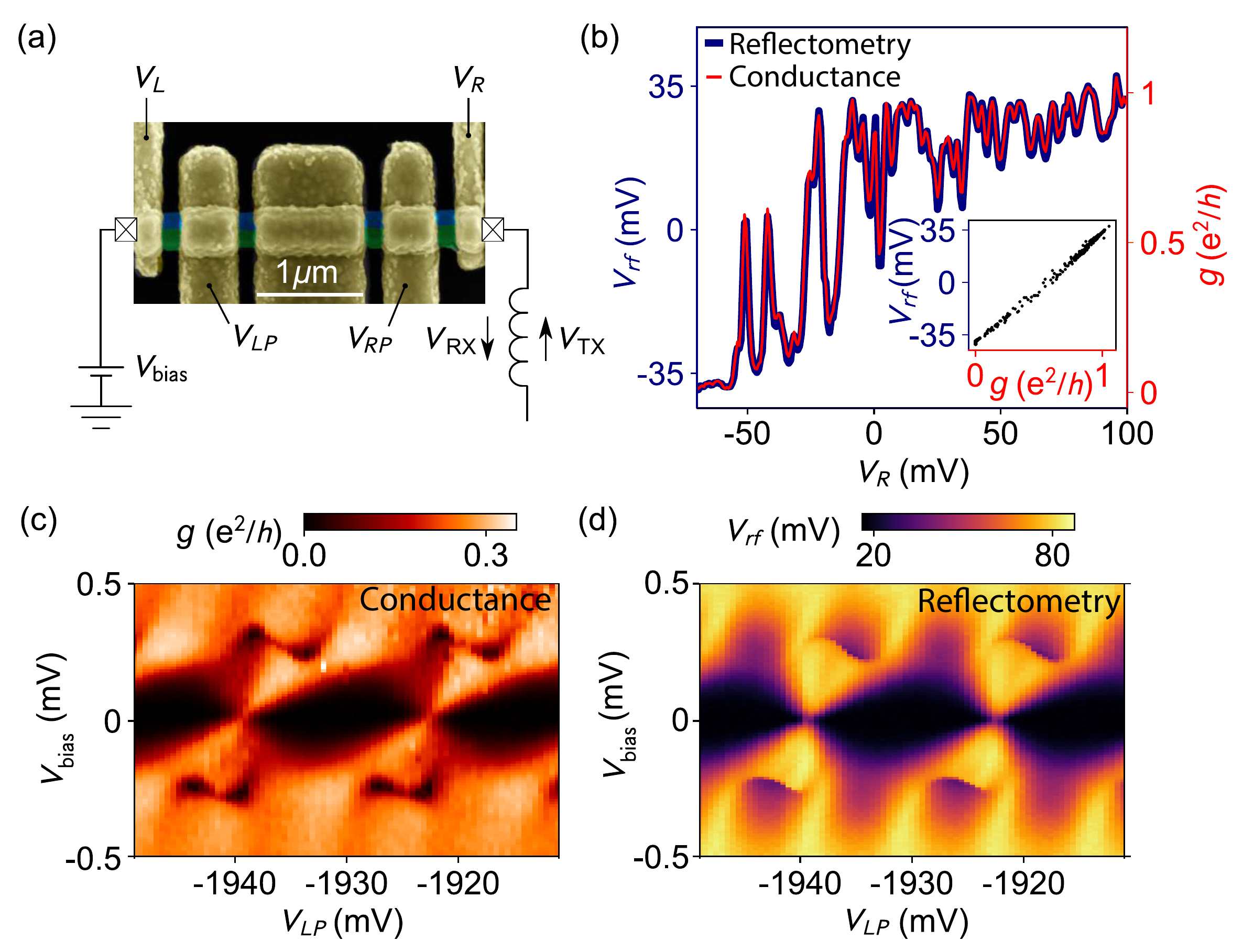}
	\caption{\textbf{Conductance via lock-in versus lead reflectometry.} Device A. (a) Scanning electron micrograph of a top-gated InAs/Al nanowire device with the relevant gates labeled. (b) Zero bias pinch-off characteristic of the right barrier gate, $V_R$, as measured by conductance (red) and reflectometry (blue). The inset shows a parametric plot of the two traces. Coulomb blockade diamonds measured by conductance (c) and reflectometry (d). In both cases, the dependence on plunger voltage $V_{LP}$ is 1\textit{e} periodic at high bias and 2\textit{e} periodic at zero bias.}
\end{figure}

Setting both barriers into the tunneling regime using $V_L$ and $V_R$ created a Coulomb blockaded island. A two-dimensional map of Coulomb diamonds as a function of $V_\mathrm{bias}$ and left plunger gate, $V_{LP}$, is shown in Figs.~2(c,d). At finite bias, $V_\mathrm{bias}$ $\geq$ 0.2 mV, above the superconducting gap of Al, conductance oscillations with period half the zero-bias period was found, characteristic of a superconducting island. At low bias, transport is via Cooper pairs yielding 2\textit{e} periodicity; at biases above the superconducting gap, 1\textit{e}  transport is available, halving the period.

The similarity of LF lock-in and RF reflectometry data exhibited in Figs.~2(c,d) indicates that RF reflectometry yields essentially equivalent results to LF conductance, though with a dramatic reduction of data acquisition time. For instance, a two-dimensional map of $V_{L}$ vs.~$V_{LP}$ consisting of 3000 $\times$ 1500 points (Appendix \ref{appA}) required roughly 1 hour of acquisition time, including data processing. Acquiring comparable data using LF lock-in methods with a 30 ms integration time would require $1500\times3000\times30$~ms $\sim$ 38 hours to achieve comparable SNR and resolution. 

\section{Charge sensing}

Charge sensing of a Majorana island was accomplished by placing a second NW (sensor wire), without a superconducting layer, next to the hybrid-NW Majorana device, and capacitively coupling the two NWs with a floating metallic gate \cite{charge_sensing1}. Charge sensing complements conductance and is the basis of parity readout in several theoretical proposals, for instance Ref.~\cite{AasenPRX16}. The approach is similar to schemes used for spin qubit readout \cite{long_distance_charge_sensing, charge_sensing2}. In the context of topological qubits, one can generalize the idea used in spin qubits known as ``spin-to-charge conversion,'' where a well-isolated quantum variable (spin) is read out projectively by mapping the relevant qubit state onto charge and then detecting charge \cite{petta2,Vandersypen}. In a similar way, the parity of a Majorana island grounded via a trivial superconductor, a well-isolated quantum state, can be read out projectively as a charge state if the island is gated into isolation, forming a topological Coulomb island \cite{AasenPRX16}, a process we denote ``parity-to-charge conversion.''

\subsection{LF charge sensing}
A Majorana island formed from a gated segment of InAs/Al NW, with extended leads made from the same wire (Device B), is shown in Fig.~3(b). Regions with tunable carrier density and conductance, made by removing the Al shell, are aligned with electrostatic gates deposited in a subsequent lithography step. Local depletion of the charge carriers in these regions (tuned by $L$ and $R$ gates) creates two superconductor-insulator-superconductor tunnel junctions with a semiconductor-superconductor island in between. A T-shaped floating gate couples the superconducting island to the charge sensor NW, which was operated in the Coulomb blockade regime by depleting its barriers with gate voltage $V_{L}^{(S)}$ and $V_{R}^{(S)}$ (see Appendix \ref{AppD} for details).

\begin{figure}[tbh]
	\includegraphics[width=0.47\textwidth]{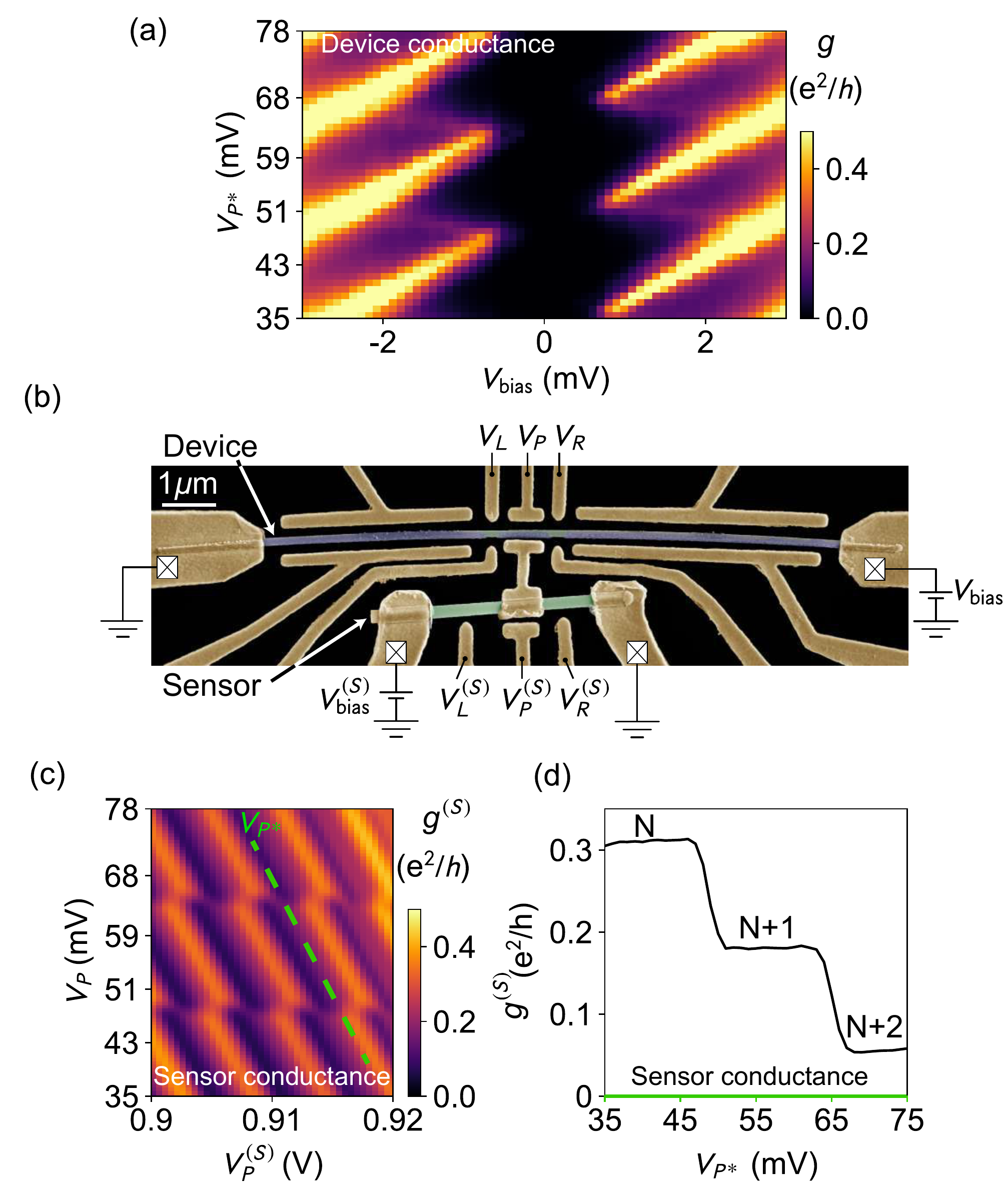}
	\caption{\textbf{Charge sensing of a superconducting island using lock-in measurement of remote charge sensor.} Device B.
		(a) Conductance through the InAs/Al nanowire device [blue in (b)] as a function of bias voltage, $V_\mathrm{bias}$, and compensated plunger voltage, $V_{P}^{*}$ (see main text for explanation of compensation). 
		(b) False color scanning electron micrograph of a device with the relevant gates labeled. A T-shaped Ti/Au floating gate couples the superconducting island (Device) to a bare InAs nanowire (Sensor). (c) Sensor conductance $g^{(S)}$ at $V_\mathrm{bias} = 0$, as a function of $V_{P}^{(S)}$ and $V_{P}$. (d) Cut along the green dashed line in (c). Distinct constant conductance value plateaus are indicated by the associated electron occupation  of the Majorana island.
	}
\end{figure}

LF lock-in  measurement of conductance through the InAs/Al NW island as a function of $V_\mathrm{bias}$ and compensated gate voltage $V_{P}^{*}$ is shown in Fig.~3(a). Compensation means that whenever the device plunger voltage $V_{P}$ is swept, the sensor plunger $V_{P}^{(S)}$ is also varied to prevent $V_{P}$ from affecting the sensor charge state via capacitive coupling, allowing the sensor to remain on a single Coulomb peak as $V_{P}^{*}$ is swept. Compensation is illustrated in Fig.~3(c), where the green dashed line shows a compensated trajectory through the space of the two plunger voltages. 

Coulomb blockade diamonds are visible in Fig.~3(a). The suppression of conductance for $|V_\mathrm{bias}|$ $<$ 0.4~mV, independent of $V_{P}^{*}$, reflects the presence of a superconducting gap in both leads, and is consistent with the gap of Al, assuming the induced gap $\Delta_I$ $\sim$ 0.2~meV is roughly equal in the three NW segments. Charging energy $E_C$ $\sim$ 0.7 meV was extracted from Coulomb diamonds of Fig.~3(a). The large charging energy, $E_C/\Delta_I > 1$ is consistent with suppressed conductance of Cooper pairs at $V_\mathrm{bias}$ = 0 \cite{Joyez1994, Matveev1993, Lotkhov2003}. The large $E_C$ results from the small capacitance between device island and the metal back-gate due to thick (500 nm) SiO$_{2}$. By comparison, Device A had 200 nm of SiO$_{2}$, reducing the charging energy to below the induced gap, leading to $2e$ Cooper-pair transport between Coulomb valleys.  

The sensor conductance, $g^{(S)}$, at zero DC bias, $V_\mathrm{bias}^{(S)}$ = 0, as a function of plunger gate voltages $V_{P}^{(S)}$ and $V_{P}$, is shown in Fig.~3(c). Conductance oscillations along the $V_{P}^{(S)}$ axis indicate that the sensor island is tuned into the Coulomb blockade regime, whereas discontinuities along $V_{P}$ reflect charge transitions in the main hybrid device. We emphasize that charge transitions are {\em not visible} in zero-bias conductance of the device [Fig.~3(a)] but {\em are visible} as plateaus in sensor conductance $g^{(S)}$ as the device charge changes by two between adjacent Coulomb valleys [Figs.~3(c,d)]. 

\subsection{RF charge sensing}
\begin{figure}[tbh]
	\includegraphics[width=0.49\textwidth]{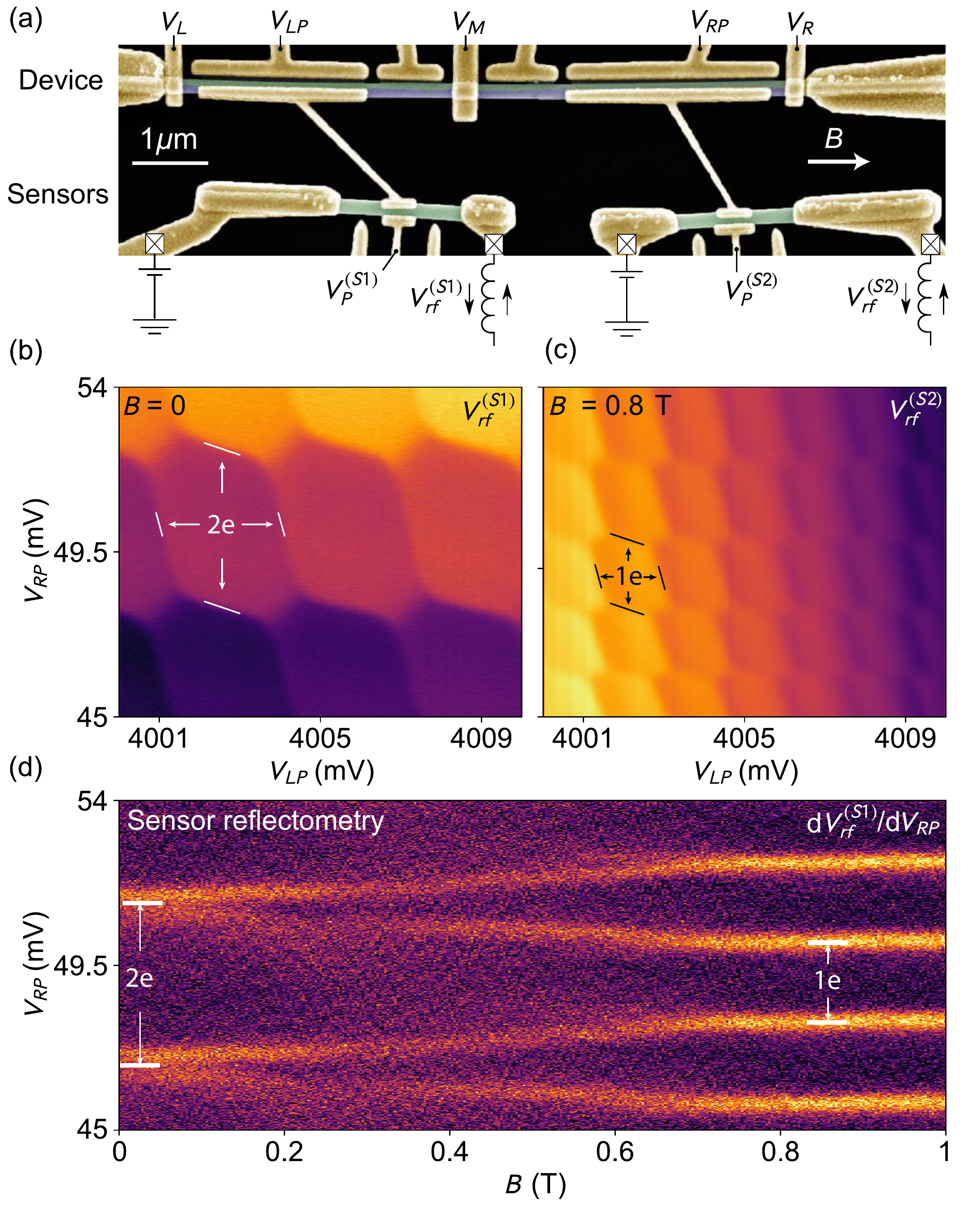}
	\caption{\textbf{RF charge sensing of double Majorana island.} Device C. (a) Scanning electron micrograph of the device measured. Voltage tunable tunnel barriers are labeled as $V_{L}$, $V_{M}$ and $V_{R}$. Island plunger gates are labeled as $V_{LP}$ and $V_{RP}$ for the left and right island respectively. (b) 2\textit{e}-2\textit{e} periodic superconducting double-island charge stability diagram measured at $B$ = 0 by RF charge sensing with a right sensor. (c) 1\textit{e}-1\textit{e} periodic double-island charge stability diagram measured at $B$ = 0.8 T with a left sensor. (d) Charge occupancy of the right island (controlled by $V_{RP}$) evolution as a function of $B$. The color map shows the measured RF demodulated signal from the right sensor ($V^{(S2)}_{rf}$) and is differentiated along the $V_{RP}$ axis. Periodicity change from 2\textit{e} to 1\textit{e} in $V_{RP}$ direction is observed as \textit{B} is increased.}
\end{figure}

A double-Majorana-island device motived by Ref.~\cite{AasenPRX16} (Device C) is shown in Fig.~4(a). Near the main device, two bare InAs NWs, capacitively coupled to each of the islands via floating gates, serve as independent charge sensors of the two islands. Each sensor is part of an independent RF circuit, with $L_{1}$ = 3.3 $\mu$H ($f_{\rm res}$ $\sim$ 60 MHz) and $L_{2}$ = 4.7~$\mu$H ($f_{\rm res}$ $\sim$ 40~MHz). Data acquisition used method (3), described above. Gates $V_{L}$, $V_{M}$, and $V_{R}$ were each set to the tunneling regime. Voltages applied to plunger gates $LP$ and $RP$ affect both the carrier density in the semiconductor and the charge offset of each island (see Appendix \ref{AppD}). Figure~4(b) shows the charge sensing signal of a 2\textit{e}-2\textit{e} periodic superconducting double-island at $B$ = 0, measured using the right charge sensor (S2), with a plane subtracted to remove cross-coupling of the plungers to the three barrier gates, $V_{L}$, $V_{M}$ and $V_{R}$.  Periodic 1\textit{e}-1\textit{e} double-island plane-fitted data, measured using the left charge sensor (S1) at finite magnetic field ($B$ = 0.8 T) parallel to NW axis, is shown in Fig.~4(c). A hexagonal pattern, characteristic of a double-island devices, is readily seen at both zero field and $B$ = 0.8 T  [Figs.~4(b,c)]. Magnetic field $B$ evolution of the right 2\textit{e} periodic island into the 1\textit{e} periodic island regime, with the left island tuned into a Coulomb valley, is shown in Fig.~4(d). The data is differentiated along $V_{RP}$ to improve visibility of the charge transitions. 

Previous works \cite{AlbretchNature,sherman} investigated nearly 1\textit{e} periodic island charge occupancy, consistent with an emerging topological phase, using conductance. Using reflectometry and charge instead has  the advantage of not require electron transport through the  device itself. As seen from Fig.~4(d), sensing is consistent with these previous transport studies \cite{AlbretchNature}. We will not focus on peak spacing and motion here, to keep the focus on measurement methods.

\subsection{Fast charge measurement and \\ signal-to-noise ratios in 1{\em e} regime}

The signal-to-noise ratio (SNR) for detecting the transfer of a single electron between islands of the double-island device in Fig.~4(a) was investigated as a function of measurement time using the pulsed gate sequence shown in Fig.~5(a). Measurements were done in an applied axial magnetic field $B$ = 0.6 T, where the charge-stability diagram shows 1$e$-1$e$ hexagons. However, in contrast to the tuning in Fig.~4(c), $V_{L}$ and $V_{R}$ were set to isolate the double-island, with negligible coupling to the source and drain. Only inter-island transitions [white and red dashed lines in Fig.~5(a)] were measurable in this configuration. 

A cyclic pulse sequence was applied to gates $LP$ and $RP$ using an arbitrary waveform generator (Tektronix 5014c), placing the system in three configurations, Initialization (I) for 150 $\mu$s, Preparation (P) for 200 $\mu$s, and Measurement (M) for a range of times from 1~$\mu$s to 50~$\mu$s  [see Fig.~5(a) inset and Appendix \ref{AppC} for details]. The preparation position and duration were chosen to yield roughly equal populations of relaxed and exited populations, which also depended sensitively on the inter-island barrier gate voltage, $V_{M}$. Results of the measurement, integrated over the measurement time, were then binned to form histograms showing the distinguishability of $N$ and $N+2$ charge-difference states ($N$ = $N_{L} - N_{R}$ is the charge difference, where $N_{L}$ and $N_{R}$ are the occupancies of the left and right islands). Note that the number of cycles used to gather histogram statistics does not affect the distinguishability of the two states. More cycles yield a convergence of the histogram to a stable, smooth bimodal distribution. On the other hand, distinguishability of the two populations is affected by the duration at the measurement (M) point. We note that only during the measurement point ($M$) readout was done by triggering the waveform digitizer card [see Appendix \ref{AppC} for details].

\begin{figure}[tbh]                     
	\includegraphics[width=0.49\textwidth]{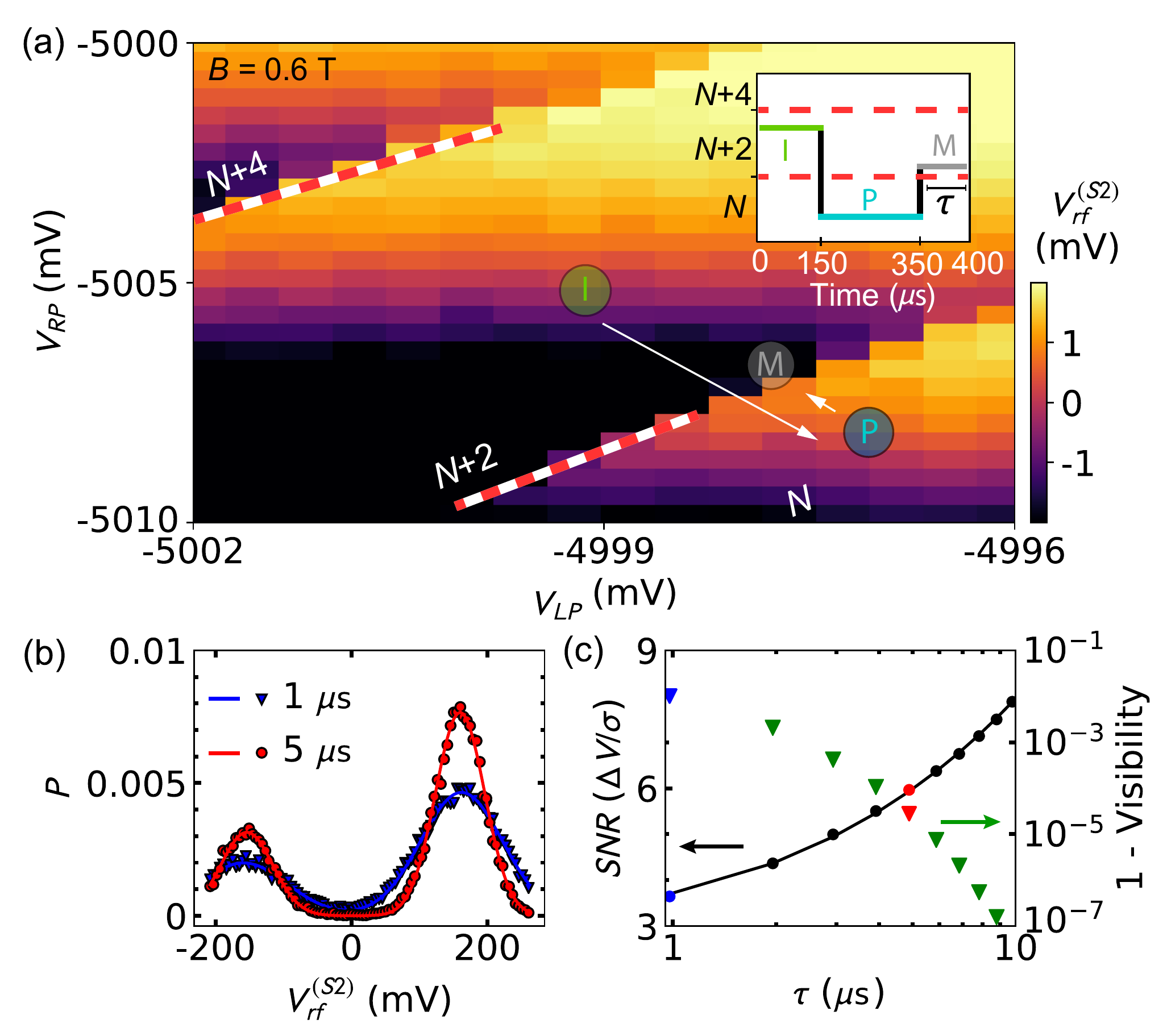}
	\caption{\textbf{Charge sensitivity and signal-to-noise ratio.} Device C. (a)  1$e$-1$e$ periodic double-island charge stability diagram at $B$~=~0.6 T measured using the right proximal charge sensor. Main device was configured such that tunneling to the left/right lead reservoirs is negligible. Only the inter-island transitions are visible with red and white dashed lines. The relative charge occupancy of the islands is marked as $N$, $N$+2 and $N$+4. The pulse sequence used to characterize signal-to-noise ratio is shown in the inset (see main text and Appendix \ref{AppC} for description), with positions I, M and P indicated on the charge stability diagram that pulsed gates $LP$ and $RP$ move the system to different gate space positions for a given amount of time. (b) Probability of single shot readout outcomes ($P$) of demodulated voltage signal $V^{(S2)}_{rf}$ for two different measurement times: $\tau$~=~1 $\mu$s (blue) and $\tau$~=~5 $\mu$s (red), at the measurement point (M) showing a bimodal relative charge state distribution. c) Signal-to-noise ratio (left axis) at the measurement point (M) together with theory fit. Extracted visibility (right axis) from the double gaussian fits (see main text) as a function of measurement time.}
\end{figure}

The resulting histogram after $10^{8}$ cycles was fit with a sum of two gaussians,

\begin{equation}
\label{gaussians}
A_{N}e^{-(V_{\rm rf}^{(S2)}-\mu_{N})^{2}/2\sigma_{N}^{2}} +A_{N+2}e^{-(V_{\rm rf}^{(S2)}-\mu_{N+2})^{2}/2\sigma_{N+2}^{2}},
\end{equation}

where $A$, $\mu$, $\sigma$ are the amplitudes, means, and standard deviations of the $N$ and $N+2$ charge differences. Measured distributions and best fits to Eq.~\eqref{gaussians} for measurement times $\tau = 1\,\mu$s and $\tau = 5\,\mu$s are shown in the Fig.~5(b). Separation of the two peaks, $\Delta$$V$, reflects the sensitivity of the charge sensor, while peak widths $\sigma_{N}$ and $\sigma_{N+2}$ result from measurement noise. We define ${\rm SNR}=~\Delta$$V$/$\sigma$, where $\sigma^{2}$~=~$\sigma_{N}^2+\sigma_{N+2}^2$. Note that Eq.~\eqref{gaussians} does not include relaxation from $N$ to $N$+2 during the measurement. A more complicated form that includes relaxation during measurement was investigated in Ref.~\cite{spin2}. In the present case, where $\tau$ is much shorter than the charge relaxation time, as set by $V_{M}$, Eq.~\eqref{gaussians} is valid. The measured SNR as a function of measurement time $\tau$ is shown in Fig.~5(c) (left axis). SNR $>$ 3 with an integration time of 1 $\mu$s was achieved. 

Figure 5(c) shows that SNR increased with measurement time, $\tau$, as expected. The simplest model of this dependence, assuming uncorrelated noise \cite{spin1}, is ${\rm SNR}(\tau)$~=~$[\Delta V/\sigma(1\,\mu{\rm s)}][(\tau+\tau_{0})/1\,\mu{\rm s}]^{1/2}$. By using fit parameter $\Delta V$~=~$175.3$~mV, $\tau_{0}$~=~$1.5$~$\mu$s and $\sigma(1\,\mu{\rm s)}$~=~$74.8$~mV, the model yields the curve shown in Fig.~5(c), which compares well with the experimentally measured ${\rm SNR}(\tau)$ in the range 1 - 10~$\mu$s. Another quantity that characterizes the quality of detection is the visibility, $V$, defined as the probability of correctly identifying excited and ground states ($N$ and $N+2$) and is expressed as $V$~=~$F_{N} + F_{N+2} - 1$, where $F_{N}$ and $F_{N+2}$ are the fidelities calculated following \cite{spin2} (see Appendix \ref{AppC} for details). The resulting dependence of visibility on measurement time, $V(\tau)$, is shown in Fig.~5(c), where again effect of relaxation during measurement are neglected. We find $V(1\,\mu{\rm s}) = 0.998$. These results are comparable or better than previously reported charge detection studies \cite{gatesens,sens1,sens2,sens3,sens4}.

\section{Conclusions}

In summary, we have investigated RF charge sensing and readout of various InAs/Al nanowire devices relevant for Majorana qubits. Two readout types were studied: First, resonant circuits were directly coupled to the device lead, yielding an improvement in measurement time by a factor of 40 compared to conventional lock-in measurements. Second, charge sensing via a second nanowire capacitively coupled via floating gate to the device allowed charge occupancy in the device to read-out non-invasively and even when visible transport is suppressed through the device. As an application, we followed the evolution of Coulomb charging from 2\textit{e} periodicity to 1\textit{e} periodicity as an axial magnetic field was increased from 0 to 0.6 T, complementing previous conductance measurement of Majorana signatures, without needing to run current through the device. Sensor quality as a function of measurement time was investigated using a pulse sequence that cycled the charge occupancies of the islands. Signal to noise ratio exceeding 3 can be achieved for integration times of 1 $\mu$s with visibility $V = 99.8$\%. Presented results show that rf resonant circuits, both directly coupled to the device, or to proximal capacitive sensors, can be used for fast and detailed characterization that conventional low-frequency techniques are not able to provide. 

\section{Acknowledgments}

We thank Shivendra Upadhyay for help with fabrication, and Wolfgang Pfaff and David Reilly for valuable discussion. Research is supported by Microsoft, the Danish National Research Foundation and by the Australian Research Council Centre of Excellence for Engineered Quantum Systems (project ID CE170100009). PK acknowledges support from ERC starting grant no.~716655. CMM acknowledges support from the Villum Foundation.
\appendix

\section{Lead reflectometry}
\label{appA}
Figure~6(a) shows a 2D map of $V_{\rm rf}$ as a function of $V_{L}$ and $V_{P}$ with 3000 $\times$ 1500 = 4.5 million points. The effective time constant per point $\tau$ = 200 $\mu$s. The data was acquired over $\sim 1$~h using RF lead reflectometry.. The estimated time of completion is about 40 hours using lock-in techniques with 30 ms integration time. Such gate maps at the moment are necessary for locating the topological regime in NW devices and as a result any technique that can speed up acquisition of such data sets can give a big advantage in experimental research.

\begin{figure}[tbh]
    \includegraphics[width=0.48\textwidth]{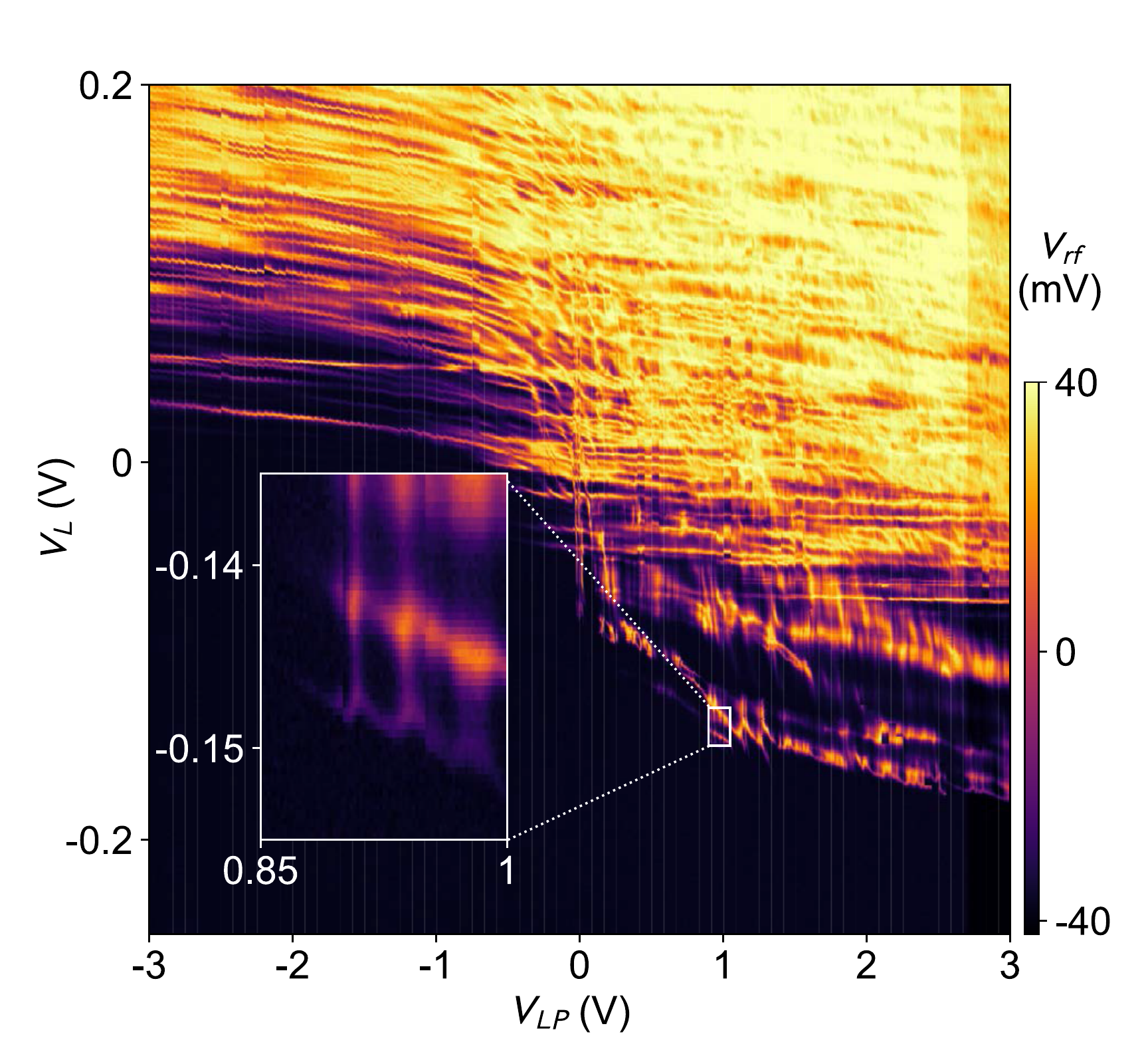}
      \caption{\textbf{Fast high-resolution charge sensing measurement using lead reflectometry.} Gate voltage map of left cutter, $V_{L}$, versus plunger gate, $V_{P}$, acquired with the lead sensing method in 1~h. The estimated time of to complete a 2D gate-gate measurement with comparable resolution using a conventional lock-in with 30 ms time constant would be $\sim 40$~h.}
\end{figure}

\section{Instruments}
\label{appB}

Reflectometry measurements presented in Fig.~2 and Fig.~6 were performed with the customised demodulation circuit presented in Fig.~7. Below we list other electronic equipment used in the experiments.

\begin{figure}[tbh]
    \includegraphics[width=0.48\textwidth]{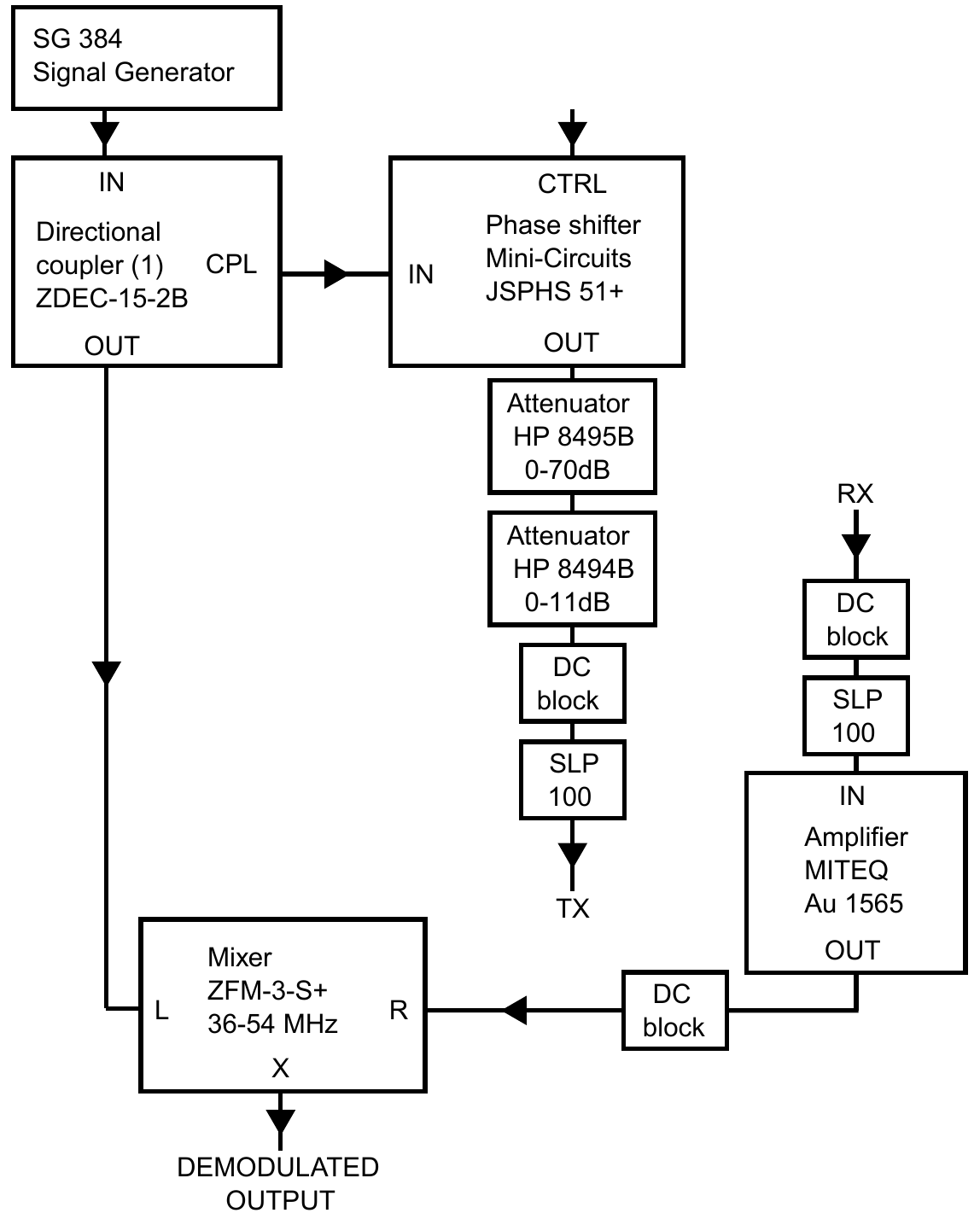}
      \caption{\textbf{Block diagram of demodulation circuit.}}
\end{figure}

\begin{enumerate}

\item Demodulation unit used for reflectometry measurements in Fig.~4 and Fig.~5: Zurich Instruments, Ultrafast Lock-in Amplifier (600 MHz) \cite{zi}

\item Current-to-voltage converter: University of Basel, Electronics Lab, Low Noise/High stability I/V converter, SP 983 with IF3602

\item Voltage sources: 48-channel QDAC, custom digital-to-analog converters, QDevil ApS \cite{qdevil}

\item Lock-in: Stanford Research SR830 DSP Lock-in amplifier

\item Waveform generator: Keysight 33500B

\item Arbitrary waveform generator: Tektronix 5014 C, 1.2 GS/s

\item Vector network analyser: Rohde $\&$ Schwarz - ZVB8

\item Directional coupler: Minicircuits ZEDC-15-2B (1 MHz - 1 GHz)

\item Microwave switch Minicircuits ZASWA-2-50DR+ (DC - 5 GHz)

\item Cryogenic 4 K amplifier: Caltech Weinreb CITLF3

\item Digitizer: AlazarTech ATS9360 - 12 bit, 1.8 GS/s

\end{enumerate}
\vspace{0.5 cm}

\section{Signal-to-noise ratio and visibility}

\label{AppC}

The extraction of signal-to-noise ratio (SNR) and visibility was accomplished with the following pulse sequence cycle [Fig.~5(a) inset]. The pulse sequence starts with a fixed amplitude voltage pulse on gates $RP$ (positive voltage pulse) and $LP$ (negative voltage pulse) bringing the system to a point I for a duration of $\tau_{I}$ = 150 $\mu$s for initialization into a relative charge state $N$+2. Then, the gates $LP$ (positive voltage pulse) and $RP$ (negative voltage pulse) bring the system into a relative charge $N$ state (point P) for a time $\tau_{P}$ = 200 $\mu $s. Finally, gates $LP$ (negative voltage) and $RP$ (positive voltage) bring the system close to intra-island degeneracy point M (between $N$ and $N$+2 relative charge states) which we denote as measurement position. $V_{TX}$ excitation was controlled with microwave switch (ZASWA-2-50DR+), in order to avoid disturbances in the system during the manipulation phase ($I$ and $P$). The readout was performed only at the measurement point (M) by triggering the ATS9360, 12 bit waveform digitizer card for a total time duration of $\tau$ = 50 $\mu $s.  To build statistics $N_{\rm cycles}$ = $10^{8}$ experimental runs of the pulse sequence were performed. From histograms of $V^{(S2)}_{rf}$ measurements (with 2 mV bin size), the probability, $P_{V^{(S2)}_{rf}}$ of singe-shot outcomes can be estimated for each value of measurement time $\tau$.

For the sake of simplicity, all denoted $V_{\rm rf}$ here will refer to demodulated voltage with the right charge sensor ($V_{\rm rf}^{(S2)}$). Visibility is defined as $V$ = $F_{N}$ + $F_{N+2}$ - 1 \cite{spin2}, where $F_{N}$ and $F_{N+2}$ are the fidelities of relative charge state $N$ and $N+2$, respectively. Fidelity of a charge state $N$ is defined by $F_{N}$ = 1 - $err_{N}$, where $err_{N}$ is an error of having a pure $N$ charge state. $N+2$ state fidelity is similarly expressed as $F_{N+2}$ = 1 - $err_{N+2}$. This error is calculated by cumulative normal distribution function, which for $N+2$ state is $\int_{-\infty}^{V_{T}}n_{N+2}dV_{rf}$, where $V_{T}$ is the threshold voltage calculated by two mean Gaussian fit peak position [($\mu_{N}+\mu_{N+2})/2$], and $n_{N+2}$ is the probability density for relative charge state $N+2$ which is expressed as $e^{{(V_{\rm rf} - \mu_{N+2})^2}/2\sigma_{N+2}^{2}}/\sqrt{2\pi}\sigma_{N+2}$. Error of having a pure $N$ state can be expressed as $\int_{V_{T} }^{\infty}n_{N}dV_{rf}$, with probability density $e^{{(V_{\rm rf} - \mu_{N})^2}/2\sigma_{N}^{2}}/\sqrt{2\pi}\sigma_{N}$. Minimizing the function of two errors ($err_{N}$ and $err_{N+2}$) and then inserting found fidelities we calculate the visibility $V = 1 - err_{N} + 1 - err_{N+2} - 1$. This yields a visibility $V$~=~99.8$\%$ for an integration time of 1~$\mu$s.

\section{Fabrication}
\label{AppD}

All devices presented have nanowire (NW) diameter  $\sim$100 nm. NWs were grown using the vapor-liquid-solid technique in a molecular beam epitaxy system with the InAs [111] substrate crystal orientation \cite{Krogstrup}. Following the NW growth, Al is deposited epitaxially \textit{in situ} on several facets of the NW with an average thickness of 10 nm \cite{Krogstrup,MT1}. The NW is then positioned on a chip with a homebuilt micro-manipulator tool (Zaber xyz theta stage with Eppendorf micromanipulator (model 4r) and large working distance Leica microscope) which allows micrometer precision in placement. The Al was selectively etched using wet etchant Transene D. All patterning was performed using an Elionix ELS-7000 EBL. Next we present the details specific to fabrication of all three devices:

\begin{enumerate}

\item Device A: The InAs/Al NW has Al shell on two of its facets and is fabricated on Si chip covered with 200 nm of $\text{SiO}_{\text{2}}$. The Ti-Au contacts (5 nm + 150 nm) were evaporated after performing RF milling to remove the oxide from the NW. Then, 7 nm of $\text{HfO}_{\text{2}}$ was deposited by atomic layer deposition. Finally the last set of Ti-Au gates (5 nm + 150 nm) was evaporated.

\item Device B: The InAs/Al NW has Al shell on two of its facets and is fabricated on Si chip covered with 500 nm of $\text{SiO}_{\text{2}}$. Then first set of Ti-Au contacts (5 nm + 100 nm) were evaporated after performing RF milling to remove the oxide from the NW. Finally the last set of Ti-Au gates (5 nm + 100 nm) was evaporated.

\item Device C: The InAs/Al NW has Al shell on two of its facets and is fabricated on Si chip covered with 200 nm of $\text{SiO}_{\text{2}}$. The Ti-Au contacts (5 nm + 150 nm) were evaporated after performing RF milling to remove the oxide from the NW. Then, 5 nm of $\text{HfO}_{\text{2}}$ was deposited by atomic layer deposition. Finally the last set of Ti-Au gates (5 nm + 150 nm) was evaporated. 

\end{enumerate}


\begin{thebibliography}{70}
	
	\bibitem{AasenPRX16} D.~Aasen, M.~Hell, R.~V.~Mishmash, A.~Higginbotham, J.~Danon, M.~Leijnse, T.~S.~Jespersen, J.~A.~Folk, C.~M.~Marcus, K.~Flensberg, J.~Alicea, Milestones toward Majorana based quantum computing.~Phys.~Rev.~X \textbf{6}, 031016 (2016).
	
	\bibitem{Vijay}, S.~Vijay and L.~Fu, Teleportation-based quantum information processing with Majorana zero modes, Phys.~Rev.~B \textbf{94}, 235446 (2016).
	
	\bibitem{Plugge}
S.~Plugge, A.~Rasmussen, R.~Egger, and K.~Flensberg, Majorana box qubits. New Journal of Physics, \textbf{19}, 012001 (2017).

\bibitem{Karzig} T.~Karzig, C.~Knapp, R.~M.~Lutchyn, P.~Bonderson, M.~B.~Hastings, C.~Nayak, J.~Alicea, K.~Flensberg, S.~Plugge, Y.~Oreg, C.~M.~Marcus, M.~H.~Freedman, Scalable designs for quasiparticle-poisoning-protected topological quantum computation with Majorana zero modes, Phys.~Rev.~B \textbf{95}, 235305 (2017).
	


	\bibitem{Non-Abelian} J.~Alicea, Y.~Oreg, G.~Refael, F.~von Oppen, M.~P.~A.~Fisher, Non-abelian statistics and topological quantum information processing in 1d wire networks, Nature Physics \textbf{7}, 412 (2011).
	
	\bibitem{Alicea} J.~Alicea, New Directions in the Pursuit of Majorana fermions in solid state systems, Rep.~Prog.~Phys.~\textbf{75}, 076501 (2012).
	
	\bibitem{Reiher7555} M.~Reiher,~N.~Wiebe,~K.~Svore.~D.~Wecker,~M.~Troyer.~Elucidating reaction mechanisms on quantum computers, National Academy of Sciences, \textbf{114}, 7555--7560  (2017).
	
	\bibitem{Mourik} V.~Mourik, K.~Zuo, S.~M.~Frolov, S.~R.~Plissard, E.~P.~A.~M.~Bakkers.L.~P.~Kouwenhoven, Signatures of Majorana fermions in hybrid superconductor-semiconductor nanowire devices, Science~\textbf{354}, 1003 (2012).
	
	\bibitem{Quantized_Majorana} H.~Zhang, C-X.~Liu, S.~Gazibegovic, D.~Xu, J.~ A.~Logan, G.~Wang, N.~v.~Loo, Jouri D.~S.~Bommer, M.~W.~A.~d.~Moor, D.~Car, R.~L.~M.~O.~h.~Veld, P.~J.~v.~Veldhoven, S.~Koelling, M.~A.~Verheijen, M.~Pendharkar, D.~J.~Pennachio, B.~Shojaei, J.~S.~Lee, C.~J.~Palmstrøm, E.~P.~A.~M.~Bakkers, S.~D.~Sarma, L.~P.~Kouwenhoven, Quantized Majorana conductance, Nature \textbf{556}, 74–79  (2018).
	
	\bibitem{MT1} M.~T.~Deng, S.~Vaitiekenas, E.~B.~Hansen, J.~Danon, M.~Leijnse, K.~Flensberg, J.~Nyg\aa rd, P.~Krogstrup, C.~M.~Marcus, Majorana bound state in a coupled quantum-dot hybrid-nanowire system, Science~\textbf{354}, 1557 (2016).
	
	\bibitem{AlbretchNature} S.~M.~Albrecht, A.~P.~Higginbotham, M.~Madsen, F.~Kuemmeth, T.~S.~Jespersen, J.~Nyg\aa rd, P.~Krogstrup, C.~M.~Marcus, Exponential protection of zero modes in Majorana islands, Nature \textbf{531}, 206 (2016).
	
	\bibitem{Lossqpp} D.~Rainis,D.~Loss, Majorana qubit decoherence by quasiparticle poisoning, Phys.~Rev.~B,~\textbf{85}, 174533 (2012).
	
	\bibitem{SCq3} J.~M.~Martinis, S.~Nam, C.~Urbina, J.~Aumentado, Rabi Oscillations in a Large Josephson-Junction Qubit, Phys.~Rev.~Lett.~\textbf{95}, 11 (2002).~
	
	\bibitem{SCq2} A.~Wallraff, D.~I.~Schuster, A.~Blais, L.~Frunzio, R.~-S.~Huang, J.~Majer, S.~Kumar, S.~M.~Girvin, R.~J.~Schoelkopf, Strong coupling of a single photon to a superconducting qubit using circuit quantum electrodynamics, Nature~\textbf{89}, 162–167 (2004).
	
	\bibitem{SCq1} A.~Wallraff, D.~I.~Schuster, A.~Blais, L.~Frunzio, J.~Majer, M.~H.~Devoret, S.~M.~Girvin,R.~J.~Schoelkopf, Approaching Unit Visibility for Control of a Superconducting Qubit with Dispersive Readout, Phys.~Rev.~Lett.~\textbf{95}, 060501 (2005).~
	
	\bibitem{SCq4} L.~Casparis, T.~W.~Larsen, M.~S.~Olsen F.~Kuemmeth, P.~Krogstrup, J.~Nygård, K.~D.~Petersson, C.~M.~Marcus, Gatemon Benchmarking and Two-Qubit Operations, Phys.~Rev.~Lett.~\textbf{116}, 150505 (2016).
	
	\bibitem{Reilley1} D.~J.~Reilly, J.~M.~Taylor, J.~R.~Petta, C.~M.~Marcus, M.~P.~Hanson, A.~C.~Gossard, Suppressing Spin Qubit Dephasing by Nuclear State Preparation, Science~\textbf{321}, 817-821 (2008).~
	
	\bibitem{spin1} C.~Barthel, M.~Kjaergaard, J.~Medford, M.~Stopa, C.~M.~Marcus, M.~P.~Hanson, A.~C.~Gossard, Fast sensing of double-dot charge arrangement and spin state with a radio-frequency sensor quantum dot, Phys.~Rev.~B \textbf{81}, 161308(R) (2010).
	
	\bibitem{petta2} J.~R.~Petta, A.~C.~Johnson, J.~M.~Taylor, E.~A Laird, A.~Yacoby, M.~D.~Lukin, C.~M.~Marcus, M.~P.~Hanson, A.~C.~Gossard, Coherent Manipulation of Coupled Electron Spins in Semiconductor Quantum Dots, Science~\textbf{309}, 5744 (2005).
	
	\bibitem{Vandersypen} R.~Hanson, L.~P.~Kouwenhoven, J.~R.~Petta, S.~Tarucha, L.~M.~K.~Vandersypen, Spins in few-electron quantum dots, Rev.~Mod.~Phys.~\textbf{749}, 1217 (2007).~
	
	\bibitem{karl1} M.~Jung, M.~D.~Schroer, K.~D.~Petersson, J.~R.~Petta, Radio Frequency Charge Sensing in InAs Nanowire Double Quantum Dots, Appl.~Phys.~Lett.~\textbf{100}, 253508 (2012).
	
	\bibitem{karl2} M.~D.~Schroer, M.~Jung, K.~D.~Petersson, J.~R.~Petta, Radio Frequency Charge Parity Meter,~Phys.~Rev.~Lett.~\textbf{109},~166804,~(2012).~
	
	\bibitem{karl3} K.~D.~Petersson, C.~G.~Smith, D.~Anderson, P.~Atkinson, G.~A.~C.~Jones, D.~A.~Ritchie, Charge and Spin State Readout of a Double Quantum Dot Coupled to a Resonator, Nano Lett.~\textbf{10}, (2010).
	
	\bibitem{qpc_rf} D.~J.~Reilly, C.~M.~Marcus, M.~P.~Hanson, A.~C.~Gossard, Fast single-charge sensing with a rf quantum point contact, Appl.~Phys.~Lett.~\textbf{91}, 162101 (2007).
	
	\bibitem{impedance_matching} N.~Ares, F.~J.~Schupp, A.~Mavalankar, G.~Rogers, J.~Griffiths, G.~A.~C.~Jones, I.~Farrer, D.~A.~Ritchie, C.~G.~Smith, A.~Cottet, G.~A.~D.~Briggs, E.~A.~Laird, Phys.~Rev.~Applied,~\textbf{5}, 034011 (2016).
	
	\bibitem{chip_inductor} Electronic access: https://www.coilcraft.com.
	
	\bibitem{amp} Electronic access: https://www.physik.unibas.ch.
	
	\bibitem{zi} Electronic access: https://www.zhinst.com/products/uhfli, p.~162
	
	\bibitem{long_distance_charge_sensing} L.~Trifunovic, O.~Dial, M.~Trif, J.~R.~Wootton, R.~Abebe, A.~Yacoby, D.~Loss, Long-Distance Spin-Spin Coupling via Floating Gates, Phys. Rev. X.~\textbf{2}, 011006, (2012).
	
	\bibitem{charge_sensing1} Y.~Hu, H.~O.~H.~Churchill, D.~J.~Reilly, J.~Xiang, C.~M.~Lieber, C.~M.~Marcus, A Ge/Si heterostructure nanowire-based double quantum dot with integrated charge sensor, Nature Nanotechnology,~\textbf{2}, 622–625 (2007).
	
	\bibitem{charge_sensing2} G.~Tosi, F.~A.~Mohiyaddin, V.~Schmitt, S.~Tenberg, R.~Rahman, G.~Klimeck, A.~Morello, Silicon quantum processor with robust long-distance qubit couplings, Nature Communications, \textbf{8}, 450 (2017). 
	
	\bibitem{petta} J.~Stehlik, Y.-Y.~Liu, C.~M.~Quintana, C.~Eichler, T.~R.~Hartke, J.~R.~Petta, Fast Charge Sensing of a Cavity-Coupled Double Quantum Dot Using a Josephson Parametric Amplifier, Phys.~Rev.~Applied,~\textbf{4}, 014018 (2015).
	
	\bibitem{Joyez1994}P.~Joyez, P.~Lafarge, A.~Filipe, D.~Esteve, M.~H.~Devoret, Observation of parity-induced suppression of Josephson tunneling in the superconducting single electron transistor, Phys.~Rev.~Lett.~\textbf{72}, 2458 (1994).
	
	\bibitem{Matveev1993}K.~A.~Matveev, M.~Gisself{\"{a}}lt, L.~I.~Glazman, M.~Jonson, R.~I.~Shekhter, Parity-induced suppression of the Coulomb blockade of Josephson tunneling, Phys.~Rev.~Lett.~\textbf{70}, 2940 (1993).
	
	\bibitem{Lotkhov2003}S.~V.~Lotkhov, S.~A.~Bogoslovsky, A.~B.~Zorin, J.~Niemeyer, Cooper pair cotunneling in single charge transistors with dissipative electromagnetic environment
	Phys.~Rev.~Lett.~\textbf{91}, 197002 (2003).
	
	
	\bibitem{sherman} D.~Sherman, J.~S.~Yodh, S.~M.~Albrecht, J.~Nyg\aa rd, P.~Krogstrup, C.~M.~Marcus, Normal, superconducting and topological regimes of hybrid double quantum dots, Nature Nano.~\textbf{12} (2017).
	
	\bibitem{spin2} C.~Barthel, D.~J.~Reilly, C.~M.~Marcus, M.~P.~Hanson, A.~C.~Gossard, Rapid single-shot measurement of a singlet-triplet qubit, Phys.~Rev.~Lett.~\textbf{103}, 160503 (2009).
	
	\bibitem{gatesens} D.~de~Jong, J.~van~Veen, L.~Binci, A.~Singh, P.~Krogstrup, L.~P.~Kouwenhoven, W.~Pfaff, J.~D.~Watson, Rapid detection of coherent tunneling in an InAs nanowire quantum dot through dispersive gate sensing, arXiv:1812.08609, (2018).
	
	\bibitem{sens1} M.~G.~House, I.~Bartlett, P.~Pakkiam, M.~Koch, E.~Peretz, J.~van der Heijden, T.~Kobayashi, S.~Rogge, M.~Y.~Simmons, High-Sensitivity Charge Detection with a Single-Lead Quantum Dot for Scalable Quantum Computation, Phys.~Rev.~Applied \textbf{6}, 044016 (2016).
	
	\bibitem{sens2} M.~J.~Biercuk, D.~J.~Reilly, T.~M.~Buehler, V.~C.~Chan, J.~M.~Chow, R.~G.~Clark, C.~M.~Marcus, Charge sensing in carbon-nanotube quantum dots on microsecond timescales, Phys.~Rev.~B \textbf{73}, 201402(R) (2006).
	
	\bibitem{sens3} S.~J.~Angus, A.~J.~Ferguson, A.~S.~Dzurak, R.~G.~Clark, A silicon radio-frequency single electron transistor, Appl.~Phys.~Lett.~\textbf{92}, 112103 (2008).
	
	\bibitem{sens4} M.~Yuan, Z.~Yang, D.~E.~Savage, M.~G.~Lagally, M.~A.~Eriksson, A.~J.~Rimberg, Charge sensing in a Si/SiGe quantum dot with a radio frequency superconducting single-electron transistor, Appl.~Phys.~Lett.~\textbf{101}, 142103 (2012).
	
	\bibitem{qdevil} Electronic access: https://www.qdevil.com.
	
	\bibitem{Krogstrup} P.~Krogstrup, N.~L.~B.~Ziino, W.~Chang, S.~M.~Albrecht, M.~H.~Madsen, E.~Johnson, J.~Ny\aa rd, C.~M.~Marcus and T.~S.~Jespersen, Epitaxy of semiconductor–superconductor nanowires, Nature Material, \textbf{14}, 400 (2015).
\end{thebibliography}
\end{document}